\documentstyle[12pt,a41,epsfig,wrapfig,rotating]{article}

\def\ZP #1 #2 #3 {{\it Z.\ Phys.}\ {\bf #1}\ (#2) #3}
\def\PR #1 #2 #3 {{\it Phys.\ Rev.}\ {\bf #1}\ (#2) #3}
\def\b{\beta}
\def\g{\gamma}

\begin{document}

~~~~~\\

\begin{flushright}
hep-ph/0004068\\
DESY--00-057\\
Freiburg--THEP 00/6\\
April 2000
\end{flushright}
\vspace{1.5cm}

\begin{center}
{\Large \bf Heavy Quark Pair Production in Polarized Photon--Photon Collisions}

\vspace*{2cm}

\renewcommand{\thefootnote}{\fnsymbol{footnote}}
\setcounter{footnote}{1}
{\large G.~Jikia\footnote{On leave of absence from IHEP, 142284 Protvino 
(Moscow Region), Russian Federation}}\\
\vspace*{0.5cm}
{\it Fakult\"at f\"ur Physik,\\
	Albert--Ludwigs--Universit\"at Freiburg,\\
        Hermann--Herder--Str.\ 3, 
        D--79104 Freiburg, Germany }\\
\vspace*{0.8cm}

{\large  A.~ Tkabladze}\\
\vspace*{0.5cm}
{\it DESY Zeuthen, D-15738, Zeuthen,  Germany}

\setcounter{footnote}{0}

\vspace*{3.5cm}

\end{center}

\begin{abstract}
{\small We present the next-to-leading-order cross sections of the
heavy quark-antiquark pair production in polarized photon-photon
collision for the general case of photon polarizations.  The numerical
results for top-antitop production cross sections together with
production asymmetries are obtained for linearly polarized
photon-photon collisions, including one-loop QCD radiative corrections.

}
\end{abstract}

\vspace*{1cm}

\newpage

\section{Introduction}

The experimental discovery of the Higgs boson is crucial for the
understanding of the mechanism of electroweak symmetry breaking. The
search for Higgs particles is one of the main goals for the LEP2 and
Tevatron experiments and is  one of the major motivations 
for the future Large Hadron Collider (LHC) and Linear $e^+e^-$ Collider (LC). 
Once the Higgs boson is discovered, it will be of primary importance to
determine in a model independent way its tree-level and one-loop
induced couplings, spin, parity, $CP$-nature, and its total width. In
this respect the $\gamma\gamma$ Compton Collider \cite{PLC} option of  the LC
offers a unique opportunity to produce both Standard Model ({\cal
SM}) Higgs boson and neutral Higgs states $h$, $H$, $A$ of the Minimal
Supersymmetric Standard Model ({\cal MSSM}) or general two Higgs
Doublet Model ({\cal 2HDM}) \cite{Hunter} as $s$-channel resonance
decaying into $b\bar b$, $WW^*$, $ZZ$ or $t\bar t$:
$$
\gamma\gamma\to h^0,\ H^0,\ A^0\to b\bar b,\ WW^*,\ ZZ,\ t\bar t.
$$
The ability to control the polarizations of back-scattered photons \cite{PLC}
provides a powerful means for exploring the $CP$ properties of any
single neutral Higgs boson that can be produced with reasonable rate
at the Photon Linear Collider \cite{CP}.  $CP$-even Higgs $0^{++}$
bosons $h^0$, $H^0$ couple to the combination
\begin{equation}
\vec{\varepsilon_1}\cdot \vec{\varepsilon_2}=-\frac{1}{2}
(1+\lambda_1\lambda_2),
\label{scalar}
\end{equation}
while a $CP$-odd $0^{-+}$ Higgs boson $A^0$ couples to
\begin{equation}
[\vec{\varepsilon_1}\times \vec{\varepsilon_2}]\cdot\vec{k_\gamma}=
\frac{\omega_\g}{2}i\lambda_1 (1+\lambda_1\lambda_2), 
\label{pseudoscalar}
\end{equation}
where $\vec{\varepsilon_i}$ and $\lambda_i$ are photon polarization
vectors and helicities. The first of these structures couples to
linearly polarized photons with the maximal strength if the
polarizations are parallel, the letter if the polarizations are
perpendicular.  Moreover, if the Higgs boson is a mixture of $CP$-even
and $CP$-odd states, as can occur {\it e.g.} in a general {\cal 2HDM}
with $CP$-violating neutral sector, the interference of these two
terms gives rise to  $CP$-violating asymmetries \cite{CP}.  Since
{\cal MSSM} Higgs particles $h^0$, $H^0$, $A^0$ decay predominantly
into $b\bar b$ or $t\bar t$ quark pairs depending on the mass of the
Higgs boson, the heavy quark pair background in $\gamma\gamma$
collisions has been studied in great detail. One-loop QCD corrections
were calculated for the photon helicity states corresponding to
projection of total angular momentum on beam axes $J_z=0$ and $J_z=\pm
2$ \cite{JT,Zaza}. Virtual one-loop QCD corrections for $J_z=0$ were
found to be especially large due to the double-logarithmic enhancement
factor, so that the corrections are comparable or even larger than the
Born contribution for the two-jet final topologies \cite{JT}. In order
to solve this theoretical problem leading QCD corrections for $J_z=0$
have been calculated at the two-loop level \cite{FadinKhozeMartin} and
recently these leading double-logarithmic QCD corrections were
resummed to all orders \cite{Sudakov}. The account of non-Sudakov form
factor to higher orders makes the $J_z=0$ cross section well defined
and positive definite in all regions of the phase space
\cite{Sudakov}. All these studies of the influence of QCD corrections
on heavy quark production in $\gamma\gamma$ collisions were
concentrated on circularly polarized  initial photons. However, for
the direct measurements of the parity of states of Higgs bosons,
(\ref{scalar}-\ref{pseudoscalar}), linear polarization of photon beams
is needed \cite{CP}.

In the present paper we consider the QCD corrections to
heavy quark-antiquark pair production in photon-photon collision for
the general case of initial photon polarizations. We mainly concentrate on 
QCD corrections for linearly polarized photon-photon 
collisions. 
 The production cross sections and spin asymmetries  for $t\bar t$-pair
production are calculated for linearly polarized photon collisions.
The measurement of spin asymmetries is necessary to determine 
the $CP$ parity of the Higgs boson.
In the scattering of linearly polarized photon off circularly polarized ones 
at one-loop level {\it azimuthal} asymmetries arise in the production of heavy
fermion pairs.
This is a pure quantum effect which does not exist at the Born level.
These type of asymmetries are suppressed by factor $m_Q^2/s$, $s$ is the c.m.s
energy of colliding photons, and are sizeable only  for $t\bar t$-pair 
production.

The paper is organized as follows.  In the next Section we recall
the basic definitions and consider the Born cross section of heavy
quark-antiquark pair production in polarized photon collisions.  
Calculations of virtual corrections are presented in
Section 3. The real gluon emission part is discussed in Section 4.
The numerical results for top-antitop pair production cross sections
and expected asymmetries for linearly polarized photon beams are
discussed in Section 5.

\section{Born Cross Section}

The cross section of heavy quark-antiquark pair production in
polarized photon-photon collision
\begin{equation}
\gamma(p_1)+\gamma(p_2)\to Q(p_3)+\bar Q(p_4)
\end{equation}
can be written in the most general form using the Stokes parameters which
describe the polarizations of initial photons.  The covariant density
matrix of polarized photon with arbitrary polarization can be written
in the following form
\begin{eqnarray}
\rho_{\mu\nu}^{(1,2)} & = & \frac{1}{2}\left(e^x_\mu e^x_\nu+
e^y_\mu e^y_\nu\right)
\pm\frac{\xi_1^{(1,2)}}{2}
\left(e^x_\mu e^y_\nu+e^y_\mu e^x_\nu\right)
\nonumber\\
&&\mp\frac{{\rm i} \xi_2^{(1,2)}}{2}
\left(e^x_\mu e^y_\nu-e^y_\mu e^x_\nu\right)
+\frac{\xi_3^{(1,2)}}{2}
\left(e^x_\mu e^x_\nu-e^y_\mu e^y_\nu\right).
\end{eqnarray}
Here $\xi_i^{(1,2)}$ are three Stokes parameters describing 
polarization of the photon with momentum $p_{1,2}$ and $e^x$ and $e^y$
denote ort vectors in $x$ and $y$ directions.

The momenta of the particles involved in the reaction 
in the c.m.s. of the initial photons are given by
\begin{eqnarray}
&p_1 = E (1;0,0,1), \quad p_2 = E (1;0,0,-1),& \label{momenta}\\
&p_3 = E(1;\b\sin\theta\cos\phi,\b\sin\theta\sin\phi,\b\cos\theta), 
\quad
p_4 = E(1;-\b\sin\theta\cos\phi,-\b\sin\theta\sin\phi,-\b\cos\theta)~,&
\nonumber
\end{eqnarray}
where $E=\sqrt{s}/2$ is the photon beam energy and $\beta=\sqrt{1-4
m_Q^2/s}$ is the quark (antiquark) velocity.  

With this definitions, the Born cross section of heavy quark pair production
in photon-photon collisions has the form
\begin{eqnarray}
     \frac{{\rm d}\sigma(\gamma\gamma \to b\bar b)}{{\rm d} t} & = & 
\frac{6 \pi\alpha^2e_Q^4}{ s^2}
\Biggl\{\left(\frac{t_1}{u_1}+\frac{u_1}{t_1}\right)
\left(1
+\tilde\xi^{(1)}_2\tilde\xi^{(2)}_2\left(1-2\frac{Y}{t_1 u_1}\right)\right) 
+ 4 \left(1+\xi^{(1)}_3+\tilde\xi^{(2)}_3\right)\frac{m_Q^2 s Y}{t_1^2 u_1^2}
\nonumber\\
&&+2\left(
\tilde\xi^{(1)}_1 \tilde\xi^{(2)}_1-\tilde\xi^{(1)}_3 \tilde\xi^{(2)}_3\right)
\biggl(1-2\frac{Y}{t_1 u_1}\biggr)
-4\tilde\xi^{(1)}_3 \tilde\xi^{(2)}_3 \frac{Y^2}{t_1^2 u_1^2}
\Biggr\}~,
\label{Born}
\end{eqnarray}
where 
$$Y=tu-m_Q^4,\quad  t_1=t-m_Q^2\quad  \mbox{and}\quad  u_1=u-m_Q^2,$$  
and $s$, $t$ and $u$ are
Mandelstam variables for the process under consideration,
$$s=(p_1+p_2)^2,\quad t=(p_1-p_3)^2,\quad u=(p_1-p_4)^2.$$
In the above formulae instead of the original Stokes parameters  their
combinations are used. The expressions of the cross sections are shorter 
and more convenient for  integration if one includes 
the dependence on the azimuthal angle in the 
Stokes parameters, $\tilde\xi^{(1,2)}_i$.
They can be expressed by original Stokes parameters 
through the equations:
\begin{eqnarray}
\tilde\xi^{(1)}_1 &=&  \xi^{(1)}_1 \cos(2\phi)-\xi^{(1)}_3 \sin(2\phi),
\nonumber\\
\tilde\xi^{(1)}_1 &=&  \xi^{(1)}_2,
\nonumber\\
\tilde\xi^{(1)}_3 &=&  \xi^{(1)}_1 \sin(2\phi)+ \xi^{(1)}_3 \cos(2\phi),
\nonumber\\
\tilde\xi^{(2)}_1 &=&  \xi^{(2)}_1 \cos(2\phi)+\xi^{(2)}_3 \sin(2\phi),
\nonumber\\
\tilde\xi^{(2)}_1 &=&  \xi^{(2)}_2,
\nonumber\\
\tilde\xi^{(2)}_3 &=& -\xi^{(2)}_1 \sin(2\phi)+ \xi^{(2)}_3 \cos(2\phi),
\end{eqnarray}
here $\phi$ is the azimuthal angle. Parameters $\tilde\xi^{(1,2)}_i$
describe photon polarization with respect to unit vectors
$\tilde{e^x}$, $\tilde{e^y}$, where $\tilde{e^x}$ lies in the reaction
plane and $\tilde{e^y}$ is orthogonal to the reaction plane.


\section{Virtual Corrections}

The first order QCD corrections to the cross section are determined by
the interference between the tree level and one-loop diagrams. At the
one-loop level no three-gluon vertex enters and the calculation is
analogous to calculations of QED corrections to Compton scattering 
for finite electron mass   (see {\it
e.g.} \cite{DennerDittmaier}).  The calculations of virtual corrections
were done by using the symbolic manipulation program FORM
\cite{FORM}. To regularise the infrared singularities we introduced an
infinitesimal mass of the gluon $\lambda$.  In the basis of the
Stokes parameters the one-loop corrections have the form
\begin{eqnarray}
\frac{{\rm d}\sigma(\gamma\gamma\to Q\bar Q)}{ {\rm d}t}  = &&
{\cal M}_0 +\tilde\xi^{(1)}_1\tilde\xi^{(2)}_1 {\cal M}_{11}
+{\rm i}(\tilde\xi^{(1)}_1\tilde\xi^{(2)}_2
+\tilde\xi^{(1)}_2\tilde\xi^{(2)}_1) {\cal M}_{12}
+\tilde\xi^{(1)}_2\tilde\xi^{(2)}_2 {\cal M}_{22}
+\tilde\xi^{(1)}_3\tilde\xi^{(2)}_3 {\cal M}_{33}
\nonumber\\
&&+(\tilde\xi^{(1)}_3+\tilde\xi^{(2)}_3) {\cal M}_{03}.
\label{Virt}
\end{eqnarray}
In addition to the Born level expression, Eq. 2, there is the new term 
proportional to the non-diagonal product 
$(\xi^{(1)}_1\xi^{(2)}_2+\xi^{(1)}_2\xi^{(2)}_1)$, which corresponds to the 
scattering of linearly polarized photon on the circularly polarized one.

The functions ${\cal M}_i$ can be expressed through scalar one-loop
integrals $B$, $C$ and $D$.
\begin{eqnarray}
      {\cal M}_0 & = &  \frac{\alpha_S \alpha^2 e_Q^4}{2\pi s}\beta\Biggl\{
2 D(s,t) \Biggl(-2 s \frac{s^2 t_1+4 m_Q^4 (s-2 t)}{t_1 u_1}+ s t- s s_1+8 m_Q^4\Biggr)
\nonumber\\
&&-4 C_1(t) \Biggl(2\frac{3 m_Q^4-s^2-m_Q^2 s}{u_1}
+2 m_Q^4 \frac{s-t_1}{t_1^2}-5 m_Q^2- s+ t\Biggr)
\nonumber\\
&&+2 B(t) \Biggl(4 m_Q^4\biggl(\frac{s+4 s_1}{s t_1 u_1}- \frac{s_0}{t_1^3}\biggr)
+4 m_Q^2\frac{3 s t-2 m_Q^2 s_1}{s t_1^2}
\nonumber\\
&&- \biggl(\frac{m_Q^2}{t}-\frac{s}{t_1}\biggr) \biggl(1-2 \frac{t_2}{u_1}\biggr)
-2 \frac{s s_1}{u_1 t_1}+1\Biggr)
\nonumber\\
&&+C_1(s) s_0 \frac{s^2+2 m_Q^2 s+2 t_1 u_1}{t_1 u_1}
-C(s) s \frac{3 s^2-2 t_1 u_1-8 m_Q^4}{t_1 u_1}
\nonumber\\
&&+4 B(s) \frac{s Y}{s_0 t_1 u_1}
+2 \ln\biggl(\frac{\lambda^2}{m_Q^2}\biggr)
\Biggl(4 \frac{m_Q^4 s^2}{t_1^2 u_1^2}- s\frac{s+4 m_Q^2}{t_1 u_1}+2\Biggr)
+16 \frac{s^2 m_Q^4}{t_1^2 u_1^2}
\nonumber\\
&&+4 \frac{m_Q^4}{t_1 u_1}
-m_Q^2 (t+u) \frac{2 m_Q^4+t_1 u_1}{t t_1 u u_1}
-4 m_Q^2 s \frac{t_1^2+u_1^2}{t_1^2 u_1^2}
-3 s \frac{6 m_Q^2+s}{t_1 u_1} 
\nonumber\\
&&+6 +(t\leftrightarrow u)\Biggr\},
\end{eqnarray}

\begin{eqnarray}
      {\cal M}_{11} &=& 
 \frac{\alpha_S \alpha^2 e_Q^4}{2\pi s}\beta\Biggl\{
2 D(s,t) \Biggl(2 s \frac{8 m_Q^4+3 s^2-7 m_Q^2 s}{u_1}
+ \frac{s^2}{Y} \biggl(s (s+m_Q^2-7 t) 
\nonumber\\
&&+2 m_Q^2 (3 t-m_Q^2)+\frac{s^3}{u_1}\biggr)
+ s t-3 m_Q^2 s+8 m_Q^4\Biggr)
\nonumber\\
&& -4 C_1(t) \Biggl(s \frac{s (s_0-5 t_1) u_1+4 m_Q^2 t_1 u_1+s^3}{u_1 Y}
+4 \frac{s^2-2 m_Q^2 s+2 m_Q^4}{u_1}-5 m_Q^2+ t\Biggr)
\nonumber\\
&&+2 B(t) \Biggl(8 \frac{m_Q^2 t}{t_1^2}-5 \frac{m_Q^2 s}{t_1 u_1}
+\frac{m_Q^2 Y}{t t_1 u_1}-2 \frac{s}{u_1}+1\Biggr)
\nonumber\\
&& +C_1(s) \Biggl(8 s^2 \frac{s-2 m_Q^2}{Y}-\frac{s^5}{t_1 u_1 Y}
-2 s \frac{3 s^2-5 m_Q^2 s+4 m_Q^4}{t_1 u_1}+2 s_0\Biggr)
\nonumber\\
&& - C(s) s \Biggl(4  \frac{s^2+6 m_Q^4-3 m_Q^2 s}{t_1 u_1}
-2 s \frac{3 s-4 m_Q^2}{Y}+\frac{ s^4}{t_1 u_1 Y}+2\Biggr)
\nonumber\\
&& +2 B(s)s  \frac{2 t_1 u_1-s^2+2 m_Q^2 s}{t_1 u_1 s_0}
 +4 \ln\biggl(\frac{\lambda^2}{m_Q^2}\biggr)\biggl(1
-2 \frac{m_Q^2 s}{t_1 u_1}\biggr)
\nonumber\\
&& +\frac{m_Q^4}{t_1 u}+\frac{m_Q^4}{t u_1}
-19 \frac{m_Q^2 s}{t_1 u_1}+6 
+(t\leftrightarrow u)\Biggr\},
\end{eqnarray}

\begin{eqnarray}
     {\cal M}_{12} & = & 
 \frac{\alpha_S \alpha^2 e_Q^4}{2\pi s}\beta\Biggl\{
 2 D(s,t) m_Q^2 s_0 \biggl(2-\frac{s s_0 t_1}{u_1 Y}\biggr)
\nonumber\\
&& -4 C_1(t) \biggl(s_0 \frac{t s-2 m_Q^2 t_1}{Y}-\frac{(s-3 m_Q^2)^2}{u_1}
-m_Q^2 \frac{m_Q^2 t_1-t s}{t_1^2}\biggr)
\nonumber\\
&&+2 B(t) m_Q^2 \biggl(\frac{ (t_1+2 m_Q^2) (5 t_1-s)}{t_1^3}
+ \frac{ ((u-2 t) t_1+2 m_Q^4}{t t_1 u_1}\biggr)
\nonumber\\
&&+C_1(s) s s_0  \biggl(\frac{s-2 m_Q^2}{t_1 u_1}-\frac{s_0}{Y}\biggr) 
 + C(s) \biggl( s_0^2 s \biggl( \frac{1}{t_1 u_1}-\frac{1}{Y}\biggr)
-8 \frac{m_Q^2 Y}{t_1 u_1}\biggr) 
\nonumber\\
&& -Y \frac{2 (t+u) m_Q^2-(t-u)^2}{ t t_1 u u_1}
+(t\leftrightarrow u)\Biggr\}~,
\end{eqnarray}

\begin{eqnarray}
      {\cal M}_{22} &=& 
 \frac{\alpha_S \alpha^2 e_Q^4}{2\pi s}\beta\Biggl\{
2 D(s,t) (2 s \frac{s^2 t+m_Q^2 s^2+4 m_Q^4 (u-t)}{t_1 u_1}
+ s (s-t-m_Q^2)-8 m_Q^4)
\nonumber\\
&&-4 C_1(t) \Biggl(2 \frac{s^2+m_Q^2 s-5 m_Q^4}{u_1}-2
 \frac{m_Q^4 (t_1-s)}{t_1^2}-t+s+5 m_Q^2\Biggr)
\nonumber\\
&&+2 B(t)\Biggl(2 m_Q^2\frac{4 t - u+3 m_Q^2}{t_1 u_1}
-4 m_Q^2\frac{3 s t- m_Q^2 t_1}{t_1^3}+3 \frac{m_Q^2 t_1-s t}{t t_1}
-2\frac{ m_Q^2 u}{t u_1}-1\Biggr) 
\nonumber\\
&&-C_1(s) \Biggl( \frac{s s_0 (s+2 m_Q^2)}{t_1 u_1}+2 s_0\Biggr)
-C(s) \Biggl(3 s \frac{s^2-8 m_Q^4}{t_1 u_1}-2 s\Biggr)
\nonumber\\
&& -4 B(s) \frac{s Y}{s_0 t_1 u_1}
+2 \ln\biggl(\frac{\lambda^2}{m_Q^2}\biggr)\Biggl( s \frac{s+4 m_Q^2}{t_1 u_1}
-2 \frac{ m_Q^2 s^3}{t_1^2 u_1^2}-2\Biggr)
\nonumber\\
&& -s \frac{8 m_Q^2 s^2-18 m_Q^2 t_1 u_1-3 t_1 s u_1}{t_1^2 u_1^2}
+m_Q^2 (u+t)\frac{2 m_Q^4+t_1 u_1}{t t_1 u u_1}-4 m_Q^4{t_1 u_1}
\nonumber\\
&&-6+(t\leftrightarrow u) \Biggr\},
\end{eqnarray}

\begin{eqnarray}
      {\cal M}_{33} & = & 
 \frac{\alpha_S \alpha^2 e_Q^4}{2\pi s}\beta\Biggl\{
-2 D(s,t) \Biggl(\frac{s^4 t_1}{u_1 Y}
+ s^2\frac{7 t s_1-4 m_Q^2 t_1-m_Q^2 (s+t)}{Y}
\nonumber\\
&&-2 s \frac{t_1 (3 s^2+m_Q^4)-7 m_Q^2 t_1 s_1-4 m_Q^4 s}{t_1 u_1}
-16 \frac{s m_Q^6}{t_1 u_1}+3 s m_Q^2-s t-8 m_Q^4\Biggr)
\nonumber\\
&&+4 C_1(t)\Biggl(\frac{s^3 t_1}{u_1 Y}-4\frac{(s-m_Q^2)^2+m_Q^4}{u_1}
+ \frac{5 s^2 t_1+4 m_Q^2 s (u_1+2 s)}{Y}
+5 m_Q^2- t\Biggr)
\nonumber\\
&&+2 B(t) \Biggl( \frac{(t+m_Q^2) (2 t+m_Q^2)}{t u_1}
-2 m_Q^2 \frac{t-5 m_Q^2}{t_1^2}
+16 m_Q^2 \biggl(\frac{u s-m_Q^4}{s t_1 u_1}+\frac{m_Q^2 t}{t_1^3}
-\frac{m_Q^4}{s t_1^2}\biggr)+3 \Biggr)
\nonumber\\
&&-C_1(s) \Biggl(2 \frac{s s_0 s_1}{t_1 u_1}+ \frac{s^3 (4 Y+s^2)}{t_1 u_1 Y}
-8 \frac{s^2 (s-2 m_Q^2)}{Y}-2 s_0\Biggr)
\nonumber\\
&&-C(s) s (2 s \frac{4 m_Q^2-3 s}{Y}+ \frac{s^4}{t_1 u_1 Y}
+4  s_1\frac{s-2 m_Q^2}{t_1 u_1}+2)
\nonumber\\
&&-2 B(s) s \frac{s^2-2 s m_Q^2-2 t_1 u_1}{s_0 u_1 t_1}
-4 \ln\biggl(\frac{\lambda^2}{m_Q^2}\biggr)
\Biggl(2\frac{m_Q^2 s Y}{t_1^2 u_1^2}-1\Biggr)
\nonumber\\
&&-m_Q^2 s \frac{11 t_1 u_1-16 s m_Q^2}{t_1^2 u_1^2}+
m_Q^4 \frac{t u_1+t_1 u}{t t_1 u u_1}+6
+(t\leftrightarrow u)\Biggr\}~,
\end{eqnarray}

\begin{eqnarray}
      {\cal M}_{03} & = & 
 \frac{\alpha_S \alpha^2 e_Q^4}{2\pi s}\beta\Biggl\{
2 D(s,t) \Biggl(\frac{s t s_0^2}{Y}-\frac{s (s-2 m_Q^2) (s t_1-8 m_Q^2 t)}{t_1 u_1}
-2 m_Q^2 s_0\Biggr)
\nonumber\\
&&-C_1(t) \Biggl(4 s_0 \frac{s t-2 t_1 m_Q^2}{Y}-4 s_1 \frac{s-7 m_Q^2}{u_1}
-4 m_Q^2 \frac{s t-t_1 m_Q^2}{t_1^2}\Biggr)
\nonumber\\
&&-2 B(t) \Biggl(18\frac{ m_Q^4}{t_1 u_1}-16 \frac{m_Q^6}{s t_1 u_1}
- m_Q^2\frac{(s-8 m_Q^2) (s t_1+2 s t-2 t_1 m_Q^2)}{s t_1^3}
\nonumber\\
&&- m_Q^2 \frac{t_1 u_1-11 t u_1-2 t t_1-t_1 m_Q^2}{t t_1 u_1}\Biggr)
-C(s) \Biggl( \frac{s s_0^2}{Y}-s \frac{s s_0-8m_Q^2 s1}{t_1 u_1}-8 m_Q^2\Biggr)
\nonumber\\
&&+C_1(s) s s_0\Biggl( \frac{s-2 m_Q^2}{t_1 u_1}-\frac{s_0}{Y}\Biggr)
+8 \ln\biggl(\frac{\lambda^2}{m_Q^2}\biggr) \frac{m_Q^2 s Y}{t_1^2 u_1^2}
\nonumber \\
&&-2 m_Q^4  \frac{8  s^2+t_1 u_1}{t_1^2 u_1^2}
 +m_Q^2 (t+u) \frac{m_Q^4+t_1 u_1}{t t_1 u u_1}
+2 m_Q^2 s \frac{t_1^2+u_1^2}{t_1^2 u_1^2}
-s \frac{s-15 m_Q^2}{t_1 u_1}
\nonumber\\
&& +4+(t\leftrightarrow u) \Biggr\}.
\end{eqnarray}
Here $t_2=t+m_Q^2$, $s_0=s-4 m_Q^2$, and $s_1=s-m_Q^2$.  
The definitions and expressions for the scalar integrals are given in the
 Appendix A of \cite{JT}.
It is worth
mentioning that all functions  ${\cal M}_i$ are expressed
through four- and three-point functions and ultraviolet finite combinations of
two-point functions.

\section{Real Gluon Emission}

The contribution of the real gluon emission to the total cross
section is separated in two parts, soft gluon emission which cancels
out the infrared divergences of virtual corrections and hard gluon
emission.  The cross section of the soft gluon emission 
can be reproduced  in a factorized form as a product of Born level 
cross section and the  infrared divergent factor:
\begin{eqnarray}
\frac{d\sigma^{soft}}{dt} = \frac{d\sigma^{tree}}{dt} R,
\end{eqnarray}
where
\begin{eqnarray}
R &=& \frac{8\alpha_s}{3\pi}\biggl\{\biggl[-1+\frac{1}{\beta}
\biggl(1-\frac{2 m_Q^2}{s}\biggr)
 \ln{\frac{1+\beta}{1-\beta} } \biggr]\ln{\frac{2k_c}{\lambda}}
\nonumber\\
&& +\frac{1}{2\beta}\biggl(1-\frac{2m_Q^2}{s}\biggr)
\biggl[{\rm Sp}\biggl(\frac{-2\beta}{1-\beta}\biggr)-{\rm Sp}\biggl(\frac{2\beta}{1+\beta}\biggr)\biggr]
\nonumber\\
&& +\frac{1}{2\beta}\ln{\frac{1+\beta}{1-\beta}}\biggr\},
\end{eqnarray}
Here $k_c$ is the soft photon energy cut and the velocity $\beta$ is
defined in Section 2.  The dependence on the gluon mass $\lambda$ is
exactly canceled while adding soft gluon emission part and virtual
corrections for any sets of Stokes parameters.
  As can be seen from
Eqs. (\ref{Born}) and (\ref{Virt}) the term proportional to the
product $\xi^{(1)}_1\xi^{(2)}_2$ appears only in the virtual part and
therefore should be infrared finite, although one-loop scalar
functions $D(s,t)$, $D(s,u)$ and $C_1(s)$ do contain infrared
logarithms. One can easily check that infrared divergent contributions 
from these scalar functions cancel each other in the ${\cal M}_{12}$-term
of Eq. 8.

The hard gluon emission part is also calculated using FORM program.
The expressions for the squared matrix elements are lengthy and we
do not reproduce them in the paper.  The integration over
three-particle phase space is done by using Monte-Carlo integration
routine VEGAS \cite{Lepage}. After adding the soft and hard gluon 
emission cross sections the final numerical results do not depend on the
imposed energy cut of emitted gluon, $k_c$.
Special care is taken to handle sharp peaks of the cross section while 
the gluon is soft or is emitted along the quark or antiquark three momenta
 (see detailed discussion in \cite{FINST}).
 These peaks correspond to the infrared and collinear
singularities in the case of massless fermions and in our case become
essential when mass of quark
is small compared to the c.m.s. energies of photons.  To arrange the infrared
and collinear singularities along the integration axis we take the
gluon energy, denominator of the quark propagator, polar angles of quark and
antiquark and azimuthal angle of reaction as integration variables.
Integration over azimuthal angle is necessary in the case of linearly
polarized photons, as in this case the cross section does depend on
the $\phi$ angle.  The infrared singularity lies on the axis of
integration over gluon energy, whereas collinear singularities are
located along the axis of quark propagator. There are also additional
peaks when quark or antiquark are produced at zero angle with respect
to the beam direction.  Such singularities are treated by
integration over the polar angles of quark and antiquark.

\section{Results and Discussion}

As was mentioned above, the helicity cross sections $\sigma(J_z=0)$,
$\sigma(J_z=2)$ for the heavy quark pair production in the circularly
polarized photon-photon collisions were considered in \cite{JT,Zaza}.
In this paper we mainly present numerical results for the production
of heavy quark-antiquark pair for the linearly polarized photons.  We
consider two cases of linear polarizations of initial photons, when
$\Delta\gamma=0$ and $\pi/2$, where $\Delta\gamma$ is the angle
between the directions of polarization vectors of the photons. The
$\Delta\gamma=0,~\pi/2)$ correspond to the collision of linearly
polarized photons with parallel and perpendicular polarizations,
respectively.  For the measurements of the Higgs boson parity it is
necessary to consider collisions of linearly polarized photons in
order to measure the polarization asymmetry \cite{CP}
\begin{equation}
A=\frac{\sigma_{\perp}-\sigma_{\|}}{\sigma_{\perp}+\sigma_{\|}}.
\end{equation}

\begin{wrapfigure}{l}{7.cm}
\vspace*{0mm} 
\centering
\epsfig{file=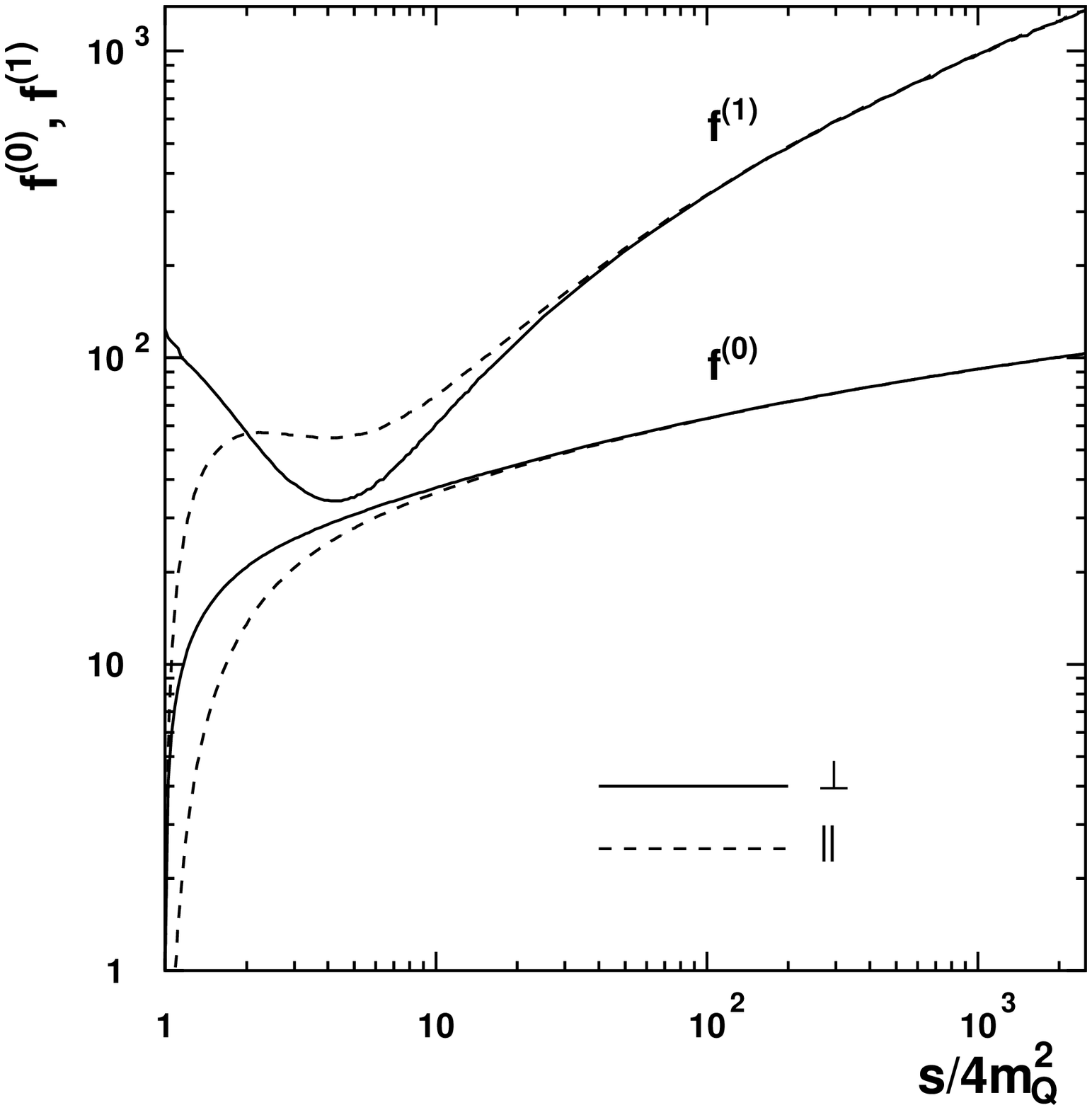,width=7.cm}
\vspace{-7mm}
\caption{\small $f^{(0,1)}_{{\perp},{\|}}$ versus $s/(4m_Q^2)$.}
\end{wrapfigure}

In fact if inclusive two-jet final states are studied then after
averaging over azimuthal angles and spins of the final particles only
three independent cross sections remain for arbitrary polarization
states of initial photons. These independent cross sections
can be taken as $\sigma_{tot}$, $\sigma(J_z=0)-\sigma(J_z=2)$ and
$\sigma_{\perp}-\sigma_{\|}$ \cite{PLC}.

For the study of the Higgs boson signal in photon-photon collisions
\cite{Borden,JT,Zaza} it was essential that the background from $b\bar
b$ quark production is suppressed by a factor of $m_b^2/s$ for $J_z=0$
at the Born level. However at the next-to-leading order the cross
section of the $b\bar b g$ production for $J_z=0$ is not suppressed
any more. Therefore experimental cuts selecting only two-jet final
states were important to suppress the $b\bar b g$ background
\cite{Borden,JT,Zaza}. In this Section we show, that the difference of
$\sigma_{\perp}-\sigma_{\|}$ is suppressed by a factor of $m_Q^2/s$
even at the next-to-leading order.

The cross section for heavy quark pair production can be cast in the
form \cite{KMS}:
\begin{eqnarray}
\sigma_{\gamma\gamma\to Q\bar Q(g)} &=& \frac{\alpha^2 Q^4 N_c}{s}
\biggl(f_{\gamma\gamma}^{(0)}+\frac{4}{3}\frac{\alpha_s}{\pi} 
f_{\gamma\gamma}^{(1)}\biggr),
\end{eqnarray}

\begin{table}
\caption{ $f^{(0,1)}$ for different polarization states of the photons
as a function of dimensionless parameter $s/(4m_Q^2)$.}
\begin{center}
\begin{tabular}{|c|c|c|c|c|c|c|}
\hline
$\frac{s}{4 m_Q^2}$ & $f^{(0)}_{\perp}$ & $f^{(0)}_{\|}$
 & $f^{(0)}_{unpol}$ & $f^{(1)}_{\perp}$ & $f^{(1)}_{\|}$ &  
$f^{(1)}_{unpol}$\\
\hline
    1 &    0.0  &    0.0 &   0.0  &  124.0 &       0.0 &   62.0 \\
    4 &   28.6  &   24.9 &  26.7  &   34.3 &      54.8 &   44.5 \\
    9 &   36.6  &   35.0 &  35.8  &   54.6 &      70.4 &   62.5	\\
   16 &   42.5  &   41.6 &  42.1  &   93.4 &     104   &   98.6	\\
   25 &   47.3  &   46.8 &  47.1  &  136   &     144   &  140	\\
  100 &   63.4  &   63.3 &  63.3  &  339   &     340   &  339	\\
  400 &   80.4  &   80.3 &  80.3  &  662   &     674   &  668	\\
 2500 &  103.2  &  103.2 & 103.2  & 1352   &     1359  & 1355	\\
\hline
\end{tabular}
\end{center}
\end{table}
where the functions $f^{(0,1)}_{\|,\perp}$ depend on the dimensionless
variable $s/(4 m_Q^2)$ only.  The numerical values of the functions
$f_{\|,\perp}^{(0,1)}$ are presented in the Figure 1 and Table I.
Because of the Sommerfeld rescattering correction, the function
$f^{(1)}_{\perp}$ is nonzero at the threshold, as one can see from the
table I.  In the both cases, $\Delta\gamma=0$ and
$\Delta\gamma=\pi/2$, the functions corresponding to QCD corrections
are positive and rising at high energies.  Taking the average of
values of these two functions one obtains the corresponding function
for the unpolarized cross section. Our results are in agreement with
previous calculations of QCD corrections for heavy quark-antiquark
production in unpolarized photon-photon collisions \cite{JT,KMS,DKZZ}.
As in the case of Born level functions $f^{(0)}_{\|}$ and
$f^{(0)}_{\perp}$, in the asymptotic regime the difference between
$f^{(1)}_{\|}$ and $f^{(1)}_{\perp}$ vanishes and each of the function
tends to the unpolarized one, $f^{1}_{unpol}$.  Such an asymptotic
behavior of the corrections can be understood considering the
helicity amplitudes for massless quarks. 
The difference of the cross sections with parallel and orthogonal polarized
photons can be expressed via the interference term of the following 
helicity amplitudes
\begin{eqnarray}
\Delta\sigma(\gamma\gamma\to Q\bar Q)=\sigma_{\perp}-\sigma_{\|}
&\simeq& {\rm Re} \sum{M^{Born}_{++}(\gamma\gamma\to Q\bar Q) 
{M^{one-loop}_{--}}^*}(\gamma\gamma\to Q\bar Q)\nonumber\\
&&+{\rm Re} \sum{M^{one-loop}_{++}(\gamma\gamma\to Q\bar Q) 
{M^{Born}_{--}}^*}(\gamma\gamma\to Q\bar Q)\\
&&+{\rm Re} \sum{M^{Born}_{++}(\gamma\gamma\to Q\bar Qg) 
{M^{Born}_{--}}^*}(\gamma\gamma\to Q\bar Qg)\nonumber
\end{eqnarray}
\begin{wrapfigure}{l}{7.cm}
\vspace*{-5mm} 
\centering
\epsfig{file=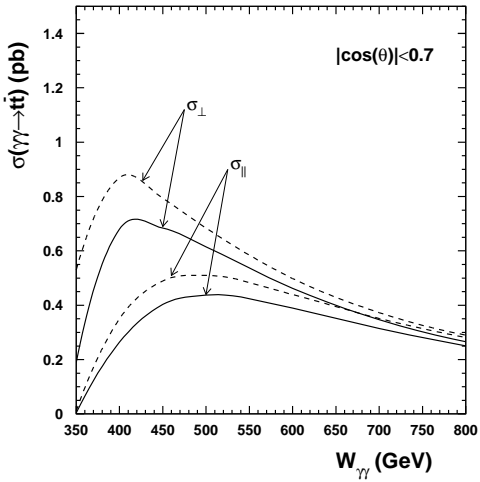,width=7.cm}
\vspace{-7mm}
\caption{\small 
The $t\bar t$ production cross section for parallel and orthogonal polarized
photon collisions versus c.m.s. energy of photons; the solid lines correspond 
to the Born cross section and the dashed lines to the cross
sections with QCD corrections.
}
\end{wrapfigure}
here the sum over the helicities of the final state particles $Q\bar
Q(g)$is implied.  The Born amplitude of the $Q\bar Q$ pair production
in the photon-photon collisions is known to vanish 
like $m_Q^2/s$ for equal photon helicities and massless quarks
\cite{GW,JT,Zaza}. Therefore first two terms in the Eq. 19 vanish in
the high energy limit. In addition, helicity amplitudes for the
process of massless $Q\bar Q$ pair production with the additional
gluon emission $\gamma\gamma\to Q\bar Q g$ identically vanish for
photon and gluon helicities $\lambda_1=\lambda_2=-\lambda_g=\pm 1$ and
arbitrary quark helicities \cite{GW}. Consequently, in the third term 
of the Eq. 19 amplitudes $M_{++}^{Born}(\gamma\gamma\to Q\bar Q g)$
and $M_{--}^{Born}(\gamma\gamma\to Q\bar Q g)$
are  nonzero only in the case when emitted gluons have different polarizations.
Therefore there is no interferention between corresponding amplitudes 
and third term of Eq. 19 term also vanishes at high energies. 
As result, the difference of the cross sections for $(Q\bar Q)$-pair
production, $\Delta\sigma(\gamma\gamma\to Q\bar Q)$  is suppressed
by a factor of $O(m_Q^2/s)$.
For $b\bar b$ production  the relative difference of the cross section 
for parallel and orthogonal polarized photons 
is less than $1\%$  for $\sqrt{s}\ge 200$ GeV, i.e. in the whole range 
of  the PLC energies.
On the other hand, for top-antitop production there is no strong suppression
of $\Delta\sigma$ by the mass of quark at energies 
$\sqrt{s}\simeq500\div800$ GeV.
The production cross sections for 
$t\bar t$-pair production are illustrated in the Figure 2
 for different helicity states of initial photons. 
The usual cut for suppression the Higgs background is imposed, 
$|\cos{\theta}|<0.7$.
The solid lines correspond to the Born level cross sections and the
dashed lines to the QCD corrected ones.
As one can see from Figure 2
the corrections are large near the threshold and decrease very rapidly with 
increasing the photon-photon c.m.s energy, $W_{\gamma\gamma}$.

In the Figure 3a the difference of two cross sections,
$\Delta\sigma(\gamma\gamma\to t\bar t)$, is shown.
%
\begin{figure}[th]
\vspace{0mm}
\centering
\begin{minipage}[c]{7.5cm}
\centering
\epsfig{file=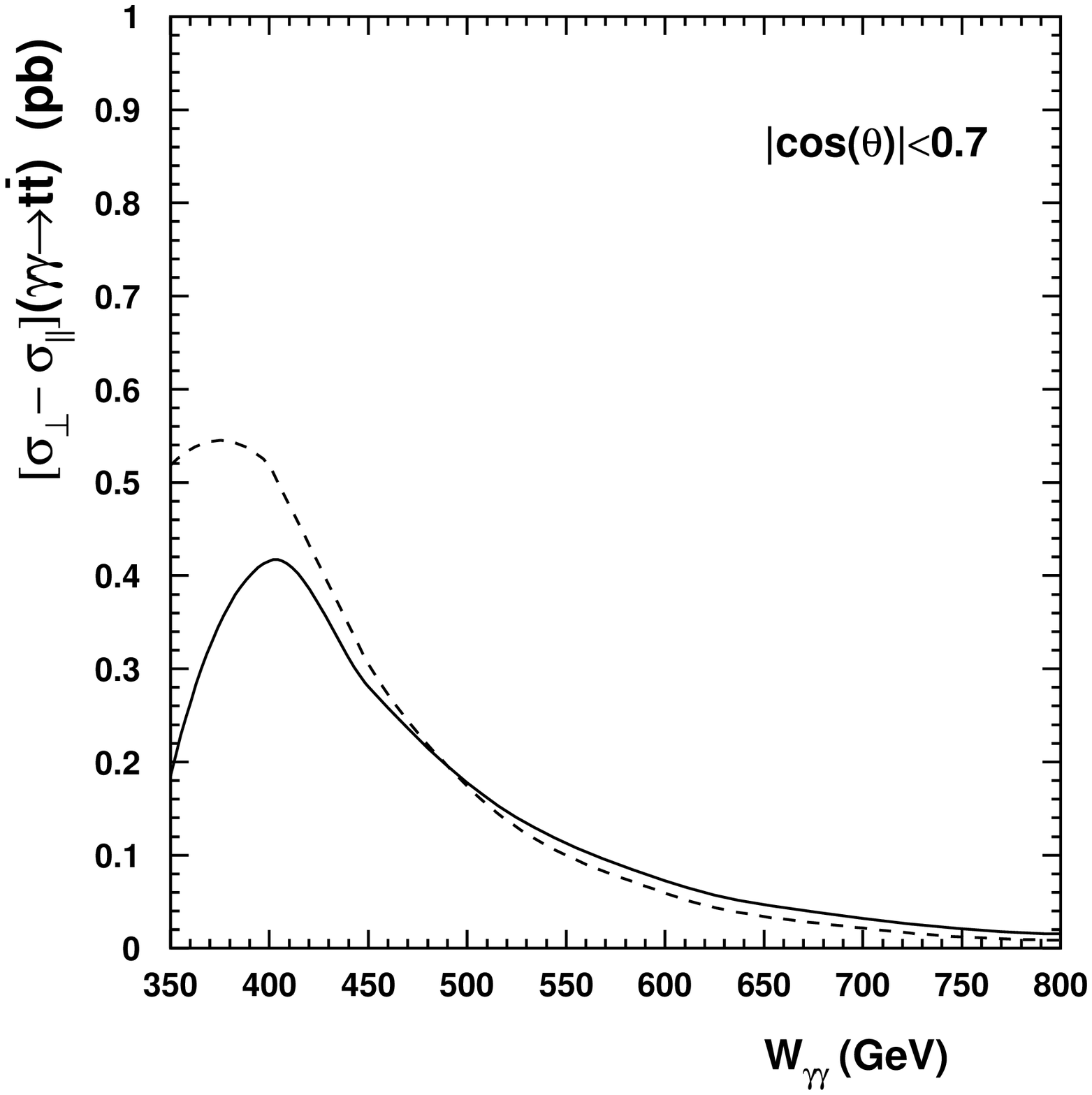,width=7.5cm}
\end{minipage}
\hspace*{0.5cm}
\begin{minipage}[c]{7.5cm}
\centering
\epsfig{file=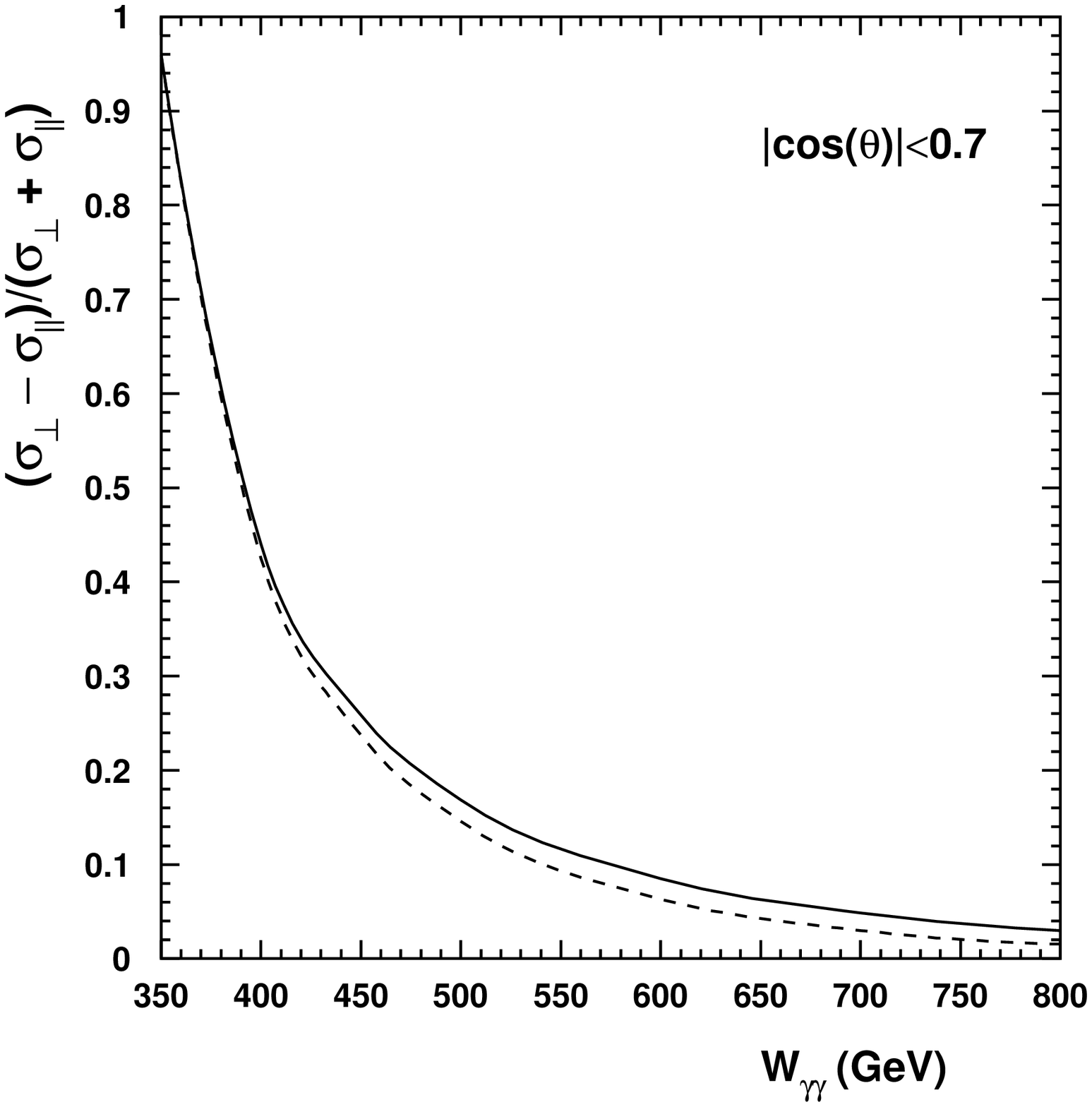,width=7.5cm}
\end{minipage}
\caption{\small
The difference of top-antitop production cross section, 
$\Delta\sigma(\gamma\gamma\to t\bar t)$; a) the absolute value of
$\Delta\sigma$, b) the relative value of $\Delta\sigma$.
The definition of lines is the same as in the previous figure.
}
\end{figure}
The correction to the $\Delta\sigma$ are rather large near the threshold,
up to $W_{\gamma\gamma}\simeq400$ GeV, and decreases rapidly.  However,
the asymmetry, $(\sigma_{\perp}-\sigma_{\|})
/(\sigma_{\perp}+\sigma_{\|})$ gets only small corrections in
the whole range of energies.

The {\it azimuthal} asymmetry of heavy quark pair  production is 
the specific effect which occurs only at one-loop level.
The relevant term in Eq. 8 proportional to imaginary unit $i$ 
corresponds to the scattering  when one photon is polarized linearly and 
the other circularly,  $\xi^{(1)}_1=\pm 1$ and
$\xi^{(2)}=\pm 1$ 
or other way round (all other Stokes parameters are zero).
This term gives
 the contributions with opposite signs to the cross
section being integrated in two different ranges of azimuthal angle,
$-\pi/2<\phi<0$ and $0<\phi<\phi/2$. 
The value $\phi=0$ corresponds to the direction
of polarization vector of linearly polarized photon.
It is obvious that the total contribution of ${\cal M}_{12}$-term to the
cross section is zero.
We define the {\it azimuthal} asymmetry 
of  $(Q\bar Q)$-pair production in the linearly polarized photon scattering off
circularly polarized one in the following way
\begin{eqnarray}
A_{\phi} = \frac{\sigma(-\frac{\pi}{2}<\phi<0)-\sigma(0<\phi<\frac{\pi}{2})}
{\sigma(-\frac{\pi}{2}<\phi<0)+\sigma(0<\phi<\frac{\pi}{2})}.
\end{eqnarray}
This  asymmetry can be  sizeable only for top-antitop production at 
PLC energies, while for the $b\bar b$-pair  production {\it azimuthal} 
asymmetries  are also  suppressed by factor  $O(m_b^2/s)$.
The expectations for the $t\bar t$-pair production asymmetry 
are shown in Figure 3. The effect is $1.5\%$ at energies of the PLC.
\begin{wrapfigure}{l}{7.cm}
\vspace*{20 mm} 
\centering
\epsfig{file=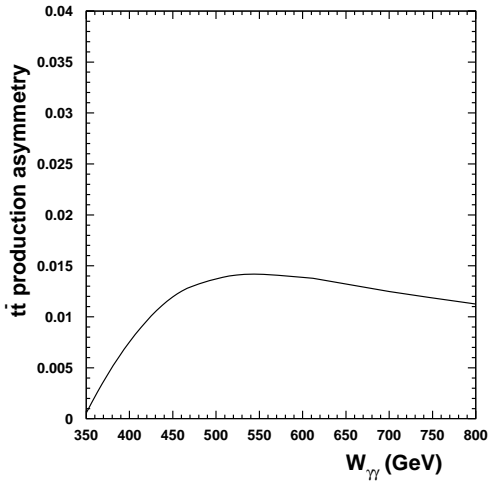,width=7.cm}
\caption{\small 
The asymmetry of top-antitop production in the photon-photon collisions.
}
\vspace*{-5mm}
\end{wrapfigure}

\section{Conclusion}

In the presented paper we derived the compact expressions for the
$\alpha_s$-corrections to the squared matrix element $\gamma\gamma\to
Q\bar Q$ for the general case of initial photon polarizations.  The
total cross sections up to the order $\alpha^2\alpha_s$ are
calculated, which are given by the sum of the tree level cross
section, contribution of the interference term between the QCD
one-loop $\alpha_s$-correction and tree level amplitude and tree level
cross sections of quark pair production accompanied by the real gluon
emission.  The numerical results are obtained for heavy quark pair
production cross sections in the case of linearly polarized photon
collisions.  The difference of the cross sections of heavy quark pair
production for parallel and perpendicular polarized photon collisions
is suppressed by factor $m_Q^2/s$.  We show that the QCD correction
for the $b\bar b$ production asymmetry is less than $1\%$ in the whole
energy range of the PLC and practically does not change the background
for the measurement of $CP$ parity of Higgs boson.  At the same time,
there is no such suppression for top-antitop production due to the
large mass of top quark.  The relevant cross sections for $t\bar
t$-pair productions are calculated.  The QCD corrections are large
near the threshold and decrease rapidly with increasing the c.m.s
energy of colliding photons.

We also calculated the {\it azimuthal} asymmetries for $t\bar t$ production.
This effect arises only at one-loop level and the asymmetry is about 
$1.5\%$ in the energy range of the PLC.

\vspace*{0.5cm}
{\bf Acknowledgments}
\vspace{0.2 cm}

We would like to thank  J.I.~Illana for  useful  discussions.

\end{document}